\documentclass[sigconf,nonacm,balance]{acmart}

\usepackage[utf8]{inputenc}

\newcommand{\zsp}{\hspace{0pt}}

\begin{document}

\title{On the critical path to implant backdoors and the effectiveness of potential mitigation techniques: Early learnings from XZ}

\author{Mario Lins}
\orcid{0000-0003-1713-3347}
\email{mario.lins@ins.jku.at}
\affiliation{%
  \institution{Johannes Kepler University Linz}
  \department{Institute of Networks and Security}
  \streetaddress{Altenberger Straße 69}
  \city{Linz}
  \country{Austria}
  \postcode{4040}
}

\author{René Mayrhofer}
\orcid{0000-0003-1566-4646}
\email{rm@ins.jku.at}
\affiliation{%
  \institution{Johannes Kepler University Linz}
  \department{Institute of Networks and Security}
  \streetaddress{Altenberger Straße 69}
  \city{Linz}
  \country{Austria}
  \postcode{4040}
}

\author{Michael Roland}
\orcid{0000-0003-4675-0539}
\email{michael.roland@ins.jku.at}
\affiliation{%
  \institution{Johannes Kepler University Linz}
  \department{Institute of Networks and Security}
  \streetaddress{Altenberger Straße 69}
  \city{Linz}
  \country{Austria}
  \postcode{4040}
}

\author{Daniel Hofer}
\orcid{0000-0003-0310-1942}
\email{dhofer@faw.jku.at}
\affiliation{%
  \institution{Johannes Kepler University Linz}
  \department{Secure and Correct Systems Lab}
  \streetaddress{Altenberger Straße 69}
  \city{Linz}
  \country{Austria}
  \postcode{4040}
}

\author{Martin Schwaighofer}
\email{martin.schwaighofer@ins.jku.at}
\affiliation{%
  \institution{Johannes Kepler University Linz}
  \department{Institute of Networks and Security}
  \streetaddress{Altenberger Straße 69}
  \city{Linz}
  \country{Austria}
  \postcode{4040}
}

\begin{abstract}
An emerging supply-chain attack due to a backdoor in XZ Utils has been identified. 
The backdoor allows an attacker to run commands remotely on vulnerable servers utilizing SSH without prior authentication.
We have started to collect available information with regards to this attack to discuss current mitigation strategies for such kinds of supply-chain attacks. 
This paper introduces the critical attack path of the XZ backdoor and provides an overview about potential mitigation techniques related to relevant stages of the attack path. 
\end{abstract}


\maketitle

\section{Introduction}
\label{sec:introduction}
On Friday, March 29th, 2024, Andres Freund ignited news papers, social media channels, and security experts around the world when he revealed~\cite{freund2024} a critical supply-chain attack due to a backdoor in the widely used XZ Utils (\href{https://nvd.nist.gov/vuln/detail/CVE-2024-3094}{CVE-2024-3094}). 
XZ Utils is a set of open source components, including xz and lzma, used for data compression.   
An attacker, using this backdoor, is able to execute malicious code with root privileges on vulnerable systems.
The seriousness of this security incident is underscored by the widespread usage of XZ Utils on a significant number of Linux and other Unix-based systems, and the CVSS rating of 10.0. 
Security experts, researchers, and other security people around the world, have already started to analyze this security incident in various aspects. 

\paragraph{Our contribution}
Based on an aggregation of the various existing analyses of the attack, this paper identifies the essential stages of the critical attack path for implanting and activating the backdoor. 
Following this path, we evaluate if, and how, various well-known techniques and best-practices could have mitigated this particular type of supply-chain attack.

\section{Preliminaries}
\label{sec:preliminaries}
This section introduces relevant preliminaries used within the critical attack path.

\subsection{XZ Utils}
XZ Utils is a widely distributed set of open source packages used for lossless data compression on Unix-like operating systems, including Linux. 
Beside supporting compression with the xz file format, XZ Utils also supports the legacy \textit{lzma} format~\cite{lzma2024}.
Because of its high compression ratio, in the last decade it has been included in many basic system components such as archival/compression tools, package installers like \texttt{dpkg} and \texttt{rpm}, the squashfs filesystem\footnote{\url{https://docs.kernel.org/filesystems/squashfs.html}}, and even the Linux kernel itself for decompression of the kernel image and initramfs\footnote{\url{https://docs.kernel.org/staging/xz.html}}.

\subsection{GNU indirect function support (IFUNC)}
The GNU indirect function support (IFUNC)~\cite{ifunc2024} is a feature of the GNU toolchain allowing developers to define multiple function implementations and to choose one of them at runtime. 
For selecting the proper implementation, the developer uses a resolver function. 
The dynamic loader calls the resolver function during startup and loads the corresponding implementation. 

\subsection{Google OSS-Fuzz}
Google's OSS-Fuzz~\cite{ossfuzz2024} is a continuous fuzzing tool for open source software to detect programming errors, like buffer overflows. 
According to the documentation~\cite{ossfuzznewproject2024} only projects with a significant user base and/or critical projects are included in the fuzzing process. 
To include a new project, it is necessary to open a pull request with information about the project (e.g.\ repository URL, primary contact, etc.)

\section{Attack Analysis}
\label{sec:attackanalysis}
This section describes relevant stages, including both human and technical aspects, that lead to the compromise of the XZ C library by incorporating a backdoor.
We start with an analysis of the currently known facts~\cite{freund2024,boehs2024,akamai2024,cox2024,coldwind2024,collin2024} around the security incident and derive a critical attack path by categorizing the available information (including the temporal sequence of the events) into individual stages.

\subsection{Critical Attack Path}
\begin{figure*}
  \centering
  \includegraphics[width=\textwidth]{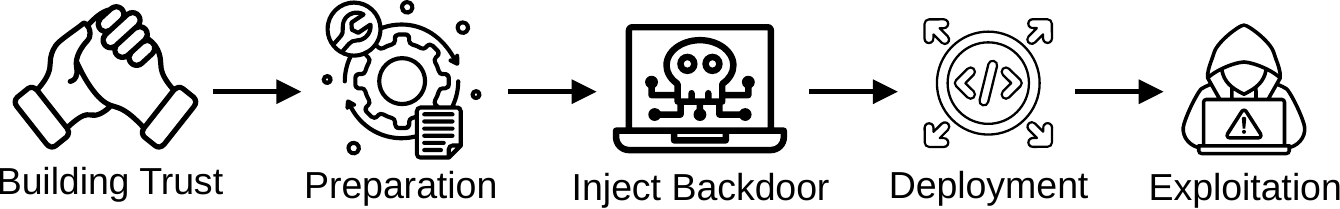}
  \caption{Critical Attack Path}
  \label{criticalattackpathfigure}
\end{figure*}
Based on various resources~\cite{freund2024,boehs2024,akamai2024,cox2024,coldwind2024,collin2024} we tried to identify the critical attack path as shown in Fig.~\ref{criticalattackpathfigure}.

\subsection{Stage 1: Building Trust}
In 2005~\cite{historylzma2024}, a small group including Lasse Collin, started to work on a project called LZMA Utils, which was later renamed XZ Utils.
In 2021, ``Jia Tan'' (presumably a pseudonym) started to support the new XZ Utils project by regularly contributing code improvements via the XZ developers mailing list. 
At the time of writing this paper, we are not aware of any malicious content in the first contributions of Jia Tan.
It appears that the intention of Jia Tan was to build up trust within the open source project, which plays a significant role with regards to the upcoming attack path. 
The first suspicious commit~\cite{prlibarchive2021} involving Jia Tan under the username ``JiaT75'' dates back to November 2021.
The commit was done within another project called ``libarchive'' and replaced the \texttt{safe\_{\zsp}fprintf} function with \texttt{fprintf}, the less secure variant of this functionality. 
This commit has been approved and merged to the main branch of that project without noticing that this might be a suspicious change. 
At the time of writing this paper, it remains uncertain whether this commit is relevant to the XZ backdoor case. 

Some months later, in April 2022, the user JiaT75 tried to submit another patch via the XZ developers mailing list~\cite{mailarchive2022}.
According to the conversation log, there was suddenly another user, called ``Jigar Kumar'' involved, who describes~\cite{mailarchive32022} this patch as ``\emph{quality of life feature}'' and complains about the slow release schedule of the project maintainer. 
Another conversation~\cite{mailarchive22022} revealed that Jigar Kumar puts additional pressure on the project maintainer Lasse Collin. 
In that conversation, Lasse Collin mentioned that the project would be an unpaid hobby and that he is already in contact with Jia Tan who ``will have a bigger role in the future''.
We detected the first commit~\cite{firstcommitasmaintainer2022} solely merged by Jia Tan from December 2022. 
Thus we assume that Jia Tan got commit permissions for the repository and potentially had been promoted to co-maintainer of the project.

\subsection{Stage 2: Preparation}
In March 2023, the primary contact mail for a component in the testing infrastructure, Google OSS-Fuzz, was changed from \texttt{{\zsp}lasse.{\zsp}collin@{\zsp}tukaani.{\zsp}org} to \texttt{{\zsp}jiat0218@{\zsp}gmail.{\zsp}com} by opening a merge request on oss-fuzz.
That change of contact was manually approved by Lasse Collin as the original maintainer, confirming that Jia Tan is an official co-maintainer of the XZ project~\cite{prprimarycontact2023}.
After some months, a new contributor called ``Hans Jansen'' provided patches to the XZ Utils project that use the GNU indirect function (IFUNC) feature~\cite{ifuncpatch12023,ifuncpatch22023}.
After some reworks and discussions, Jia Tan merged that functionality to the repository. 
Only later, it turned out, that these patches are essential for the attack because using the IFUNC feature is the function hooking mechanism eventually used by the backdoor code.
However, the use of IFUNC cannot be fuzzed by OSS-Fuzz, due to compatibility issues.
As a consequence, Jia Tan opened a pull request~\cite{oss-fuzz-pull-request-disable-ifunc} with the OSS-Fuzz project, to disable IFUNC for the fuzzed XZ Utils build. 
The OSS-Fuzz repository incorporates a GitHub-action to automatically label pull requests as ``ready to merge'' if they originate from the maintainer of a fuzzed project by verifying that the committer has a contribution history for that project and is one of the responsible project contacts. 
As both applied to Jia Tan, the pull request was labeled as \emph{ready to merge} by the bot, and then quickly approved and merged into OSS-Fuzz.

\subsection{Stage 3: Injecting Backdoor}
On Sunday, February 23th, 2024, Jia Tan added various lzma test files~\cite{backdoor2024, updatebackdoor2024} to the project.
Later, it turned out that the uploaded test files, called \texttt{bad-{\zsp}3-{\zsp}corrupt{\zsp}\_{\zsp}lzma2.{\zsp}xz} and \texttt{good-{\zsp}large\_{\zsp}com{\zsp}pressed.{\zsp}lzma} contained the binary backdoor and some other obfuscated code necessary to inject the backdoor code. 
Another essential preparation step was to add the file \texttt{build-to-host.m4} to the \texttt{.gitignore} file of the source code repository so that its source code does not get pushed to the repository. 
At this stage, the hidden backdoor is already part of the project, but is not yet deployed.

\subsection{Stage 4: Deployment}
The following day (Monday, February 24th, 2024), Jia Tan released version 5.6.0 of XZ Utils via GitHub.
In addition to the auto-generated source archives in .zip and .tar.gz form, XZ Utils releases include four additional, manually uploaded source archives with corresponding signature files.
The manually uploaded source archives, which were uploaded to GitHub for this release, also include a \texttt{build-to-host.m4} file.
The m4 file contained malicious modifications used to extract the hidden backdoor from the test files when building the deb or rpm packages. 
To prevent detection of the malicious code in the m4 file, it has not been added to source code repository, but was only part of the released source tarball.

The next step was to convince Linux distributions to include the latest, but malicious, release of XZ Utils in their package repositories. 
For that, the user Hans Jansen started to contact and pressure some Linux distributions (e.g.\ Debian~\cite{debianbug2024}).
Once the package maintainer upgraded to the new version by fetching the source code tarball, the malicious \texttt{build-to-host.m4} file gets integrated into the package build process. 
The malicious build script then initiates the steps required to extract the backdoor hidden in the test files. 
After several obfuscated modification, decompression, and execution steps on the malicious test files, an object file named \texttt{liblzma\_{\zsp}la-{\zsp}crc64\_{\zsp}fast.{\zsp}o}, which contains the backdoor code, is linked into the liblzma library via setting the respective pointer which gets resolved by the IFUNC functionality.

The following list provides a very abstract overview of the deployment stage of the backdoor: 
\begin{itemize}
	\item The malicious script \texttt{built-to-host.m4} decodes \texttt{{\zsp}bad-{\zsp}3-{\zsp}corrupt{\zsp}\_{\zsp}lzma2.{\zsp}xz} into a script.
	\item The decoded script decodes \texttt{good-large\_compressed.lzma} file to \texttt{{\zsp}liblzma\_{\zsp}la-{\zsp}crc64-{\zsp}fast.{\zsp}o}.
	\item The object file \texttt{{\zsp}liblzma\_{\zsp}la-{\zsp}crc64-{\zsp}fast.{\zsp}o} is added to the linking stage of liblzma.
\end{itemize}
More information about the malicious script can be found on~\cite{cox2024}.

\subsection{Stage 5: Exploitation}
The final stage is the exploitation stage in which the backdoor gets triggered on the infected system. 
Current investigations show that a pointer of a function used by \texttt{RSA\_{\zsp}public\_{\zsp}decrpyt} for OpenSSH (more specifically for any process of the executable located at \texttt{/usr/bin/sshd}) is changed to point on a function owned by the backdoor. 
The malicious function extracts a command from the certificate provided by the SSH client.
This is only initiated for a specific private key which makes the backdoor only available to attackers who possess the corresponding key.  
Afterwards, the command is passed to the \texttt{system()} function, which executes it.   
Thus, an attacker who controls the private SSH key can send arbitrary commands to affected SSH servers which are executed remotely without ever completing the authentication to the SSH server.

\section{Mitigation}
\label{sec:mitigations}
In this section, we speculate on some known mitigations and if/how they would potentially have impacted this particular attack path. We explicitly note that this is a preliminary analysis, and more detailed work is still ongoing.

\subsection{Organizational Security of Open Source Projects}
\label{sec:organizationalsecurity}
One of the main ideas behind open source software is to share source code so that everybody can inspect, modify, and contribute to its development. 
There are dozens of valuable open source projects that have been incorporated in other products even in the commercial sector either as library or as additional tool which enhances the functionality. 
However, some of these projects do not have a broad community contributing to the project and are mostly driven by a single enthusiast believing in that project, like XZ Utils. 
They maintain their projects over years without getting paid and mostly in their free-time, even though their code might be an essential part of larger and potentially widely used software projects.
Since the identification of the XZ backdoor, a lot of people started to rethink the organization of open source projects.
How should we deal with open source projects that do not have resources of a commercial software vendor, but are still integrated in relevant software projects?

One potential take-away from the XZ backdoor incident could be to optimize risk management processes towards open source software. 
An adapted risk management process could provide specific criteria to assess whether a particular open source library or tool can be integrated or not (cf.~\cite[Section~6.5.3]{owasp-dev-guide-oss} and \cite[PW.4.4]{nist-sp800-218}). 
This might lower the likelihood of a risk, but on the other hand could have an impact on costs as these components have to be implemented from scratch. 

A commonly used organizational strategy to mitigate risks with regards to software development is setting up a workflow to protect certain branches, like the main branch.
One specific control which could be incorporated in such a workflow could be peer reviewing so that it is not possible to merge changes to a specific branch without the approval of another authorized entity. 
GitHub, for example, provides several branch protection mechanisms, like requiring a pull request~\cite{githubbranchprotection2024} before merging any changes. 
It is also possible to additionally require a certain number of approvals before merging, which could be used to set up a peer review workflow.

This mitigation technique can also be used for other relevant stages during development, like auditing dependencies. 
An already available tool used for that particular case is called \textit{Cargo Vet}~\cite{cargovet2024} which can be used to scan third party dependencies of a project against a trusted set of audits.
Audits could be performed by the project authors or other entities trusted by them. 

Another key takeaway is the necessity of prioritizing mental health and signs of a maintainer getting overwhelmed. 
One possible approach could be that companies using open source software for earning money could support these developers in times of high load with development resources. 
This would bring the open source approach to a new level.

\subsection{User Credibility}
\label{sec:usercreditability}
By the nature of open-source projects, volunteers can and will provide their work to improve the project. 
While most contributors may act benevolently to the project, good operational security mandates to put measures in place to identify and reject malicious users and contributions. 

The credibility of traditional employees in a professional environment is governed by national laws and employment contracts.  
In exchange for an employee's workforce, they receive a reward. 
If either party of the (work) contract violates their duties, the other party can take action through the legal system. 
This may deter employees from acting maliciously; however, this protection is rather weak as an attacker will always act in a way which allows for plausible deniability. 

Therefore, we need ways to assess and improve the credibility of an anonymous user account. 
Like Laocoön is claimed to have said: ``\emph{I fear the Greeks even when bearing gifts}.''

\subsubsection{Verification of Contributors' Identity}
An important property in traditional employment is that the employer has proof of the employee's identity. 
This is a requirement for legal actions. 
However, previously this required face-to-face communication to be able to verify a physical proof of identity. 
As many open-source developers never meet~\cite{MIDHA201223}, a verification of a maintainer's physical identity was hard to achieve. 

As a counter-example, Debian package maintainers establish a physical-world web-of-trust. 
Before a GPG key is added to Debian's keyring, other maintainers are needed to sign it. 
This is preceded by a personal meeting and verification of a physical proof of identity.

In the European Union, the digital-physical world identity mapping problem is addressed by the eIDAS regulation~\cite{eidas2024}. 
It enables people in the member states to prove their identity over the internet and across borders. 
Regarding open-source development, such (or a similar) digital proof of identity could be required whenever a contributor is promoted to maintainer of a project. 
It provides a link of an otherwise anonymous online persona to a natural person. 

The downside is that digital identities can still be stolen, an identity authority could be fooled or the proof otherwise compromised. 
Also, the legitimate holder of an identity could be bribed (cf.\ section~\ref{sec:properly_fund_maintainers}) or extorted to ``lend'' their identity. 
Moreover, for certain projects, e.g.\ privacy enhancing technologies, that collide with policy in certain jurisdictions, revealing the physical-world identity of project maintainers to the public may expose them to repressions or legal actions.

\subsubsection{Strong Authentication of Contributions}
Trusted open-source maintainers are recognized by holding the credentials to their accounts. 
If attackers obtain those, they can create arbitrary trusted commits. 
The only person able to detect those rogue commits is the maintainer themselves. 

For increasing the credibility of accounts and their commits, they should be hardened against loss of credentials, for example by multi factor authentication. 
Furthermore, commits in Git can be signed using PGP~\cite{git-signing2024} which adds another layer of hardening against forging seemingly trusted commits. 

This shifts the problem towards the security of those additional factors. 
Hardware tokens, like those implementing the FIDO2 standards, are believed to provide excellent security. 
However, particularly at a stage where hardware tokens are not (yet) widely used, a well-funded adversary may be capable of distributing forged hardware tokens (mimicking legitimate ones but with e.g.\ a poisoned random number generator) to specific developer communities. 
Such tokens may then enable the attacker to efficiently calculate developers' private keys from observed public keys in order to impersonate the legitimate users. 

A protection against broken authentication mechanisms then requires additional effort in the form of checking transparency logs (cf.\ section~\ref{sec:transparencylogs}) or by peer-reviewing the code at critical times, for example before merging.

\subsubsection{Building Trust by Contributions}
In the world of blockchain, some cryptocurrencies proposed a \emph{proof-of-stake} model. 
The reasoning was that an attacker to the blockchain would require the majority of available tokens in the network and an attack would therefore be sufficiently unlikely. 

If we extend this reasoning to open-source projects, we get a model in which a contributor providing the most contributions, funds, or any other resource gains more influence on the project. 
In fact, providing valuable contributions is how users earn trust from the maintenance teams. 
In the case of the attack on XZ, user JiaT75 had become a maintainer through that approach and was, therefore, able to get a pull request related to their own project~\cite{oss-fuzz-pull-request-disable-ifunc} approved on another project (Google's OSS-Fuzz). 
While no malicious code was attached at this point, the pull request disabled checks to help avoid detection of the backdoor later. 
As user jonathanmetzman, a security engineer working for Google, pointed out~\cite{oss-fuzz-pull-request-disable-ifunc-comment}, the pull request was automatically approved as user JiaT75 was recognized as a maintainer of the XZ project. 
Furthermore, jonathanmetzman explained that it is fully up to the developers which parts of a project are tested by OSS-Fuzz.

As we can see by this example, maintainers of an open-source project are highly trusted by external entities. 
Therefore it should be harder to become a maintainer than simply investing enough resources as a group of determined attackers easily surpassing the capabilities of a single person. 

The opposite extreme is done by the Tor project. 
The network is governed by a limited number of special-purpose relays called directory authorities. 
They agree on a list of nodes considered part of the Tor network and therefore usable for routing traffic. 
Directory authorities are run by a hand-picked number of people who are personally known and trusted by the Tor community. 
In such a scheme the possible number of maintainers is limited. 
However, the selection process and the fact that a majority of directory authorities is required for a decision, ensures a very high level of credibility of the decision.

\subsubsection{Adequately Funding Maintainers}
\label{sec:properly_fund_maintainers}
A conceptually simple albeit not simple to implement solution would be to sufficiently fund open source project maintainers. 
Financially secured maintainers can invest more time in code quality as they are not required to develop their projects after their day jobs. 
Hence, they are more likely to invest time into attack preventive measures like peer-reviews. 
Furthermore, less stressed maintainers are less likely to offset security critical tasks to contributors they cannot fully trust. 
As an example, user JiaT75 was able to add themselves as the primary contact for XZ with Google's OSS-Fuzz project~\cite{oss-fuzz-pull-request-add-jiat75-to-contacts}

A big factor in user credibility is that an honest maintainer remains honest. 
As cybersecurity mostly works by increasing the cost of an attack against the potential gain, simply bribing an existing maintainer may become cheaper than forging a new one. 
Consequently, well funded and financially stable maintainers are more secure as they are more expensive to bribe.

\subsection{Enhanced Trust through Transparency Logs}
\label{sec:transparencylogs}
Another vital aspect when it comes to security of open source project is trust. 
Collaborating on a project requires a certain level of trust although everybody has the possibility to review code changes. 
Considering the process of open source development projects, we face certain areas where trust is needed to some extent.
The first appearance in the XZ case where trust was crucial, occurred in stage one, where Jia Tan claims to be a supportive programmer for the project over years.
In other cases, the relevance of the user identity became evident when modifying the primary contact related to the Google OSS Fuzz project, for instance. 

Based on these trust anchors, we could argue that one of the primary trust assets is the user identity which provides authenticity.
A common approach to verify if an asset has been produced by a specific identity is the use of asymmetric cryptography which ensures that only someone who possesses the private key can sign an artifact.
However, these private keys could be lost or stolen and thus an unauthorized person could distribute an artifact which is signed by the trusted owner.
One effective mitigation strategy to make unauthorized distribution attempts transparent and thus detectable is to use transparency logs, as already shown in similar scenarios like within distribution systems of mobile apps~\cite{2023-lins-nordsec}. 
In particular, we argue that every open source artifact that gets published should also be logged so that everyone can verify if it is properly logged and the legitimate owner can verify if there is a log available, even though the owner did not publish any artifact. 

Another mitigation technique that could be used to prevent unauthorized or hidden changes in software projects are reproducible builds~\cite{reproducibleBuildsWebsite} combined with transparency logs. 
If an artifact can be built reproducibly from a specific version of the source code, it is possible to verify if the artifact matches the source code. 
However, in case of an unauthorized distribution attempt, the source code could also be manipulated. 
To address this particular risk we suggest creating a transparency log entry including the hash of the artifact and a reference to the specific version of the source code. 
To verify the authenticity and integrity of the artifact, a verifier would first fetch the distributed artifact and the corresponding log entry. 
Afterwards, the verifier could try to build the artifact reproducibly from the source given in the transparency log. 
In case the hash value of the distributed artifact matches the hash value resulting from the reproducible build based on the source of the transparency log, the verifier can be sure that the artifact is based on the source code given in the transparency log. 
If an attacker was able to steal the signing key of the original maintainer, monitoring the log would make a distribution attempt transparent and thus detectable.

\subsection{Chain of Custody}
\label{sec:reproduciblebuilds}
This attack on XZ Utils can teach us more than one lesson about provenance and maintaining a transparent chain of custody in the path from source code to binary release packages.
Historically, source tarballs were the primary means to distribute open source software releases.
Since then, development and publication of open source software broadly shifted into public version control systems (VCS).
When a developer publishes a release on e.g.\ GitHub, a source tarball containing the exact code from the tagged version in the repository is automatically generated.
However, these release tarballs do not contain any cryptographic link to the developer (such as, for instance, a signature) and require full trust in the infrastructure of the version control platform.
In addition, developers are allowed to upload extra artifacts.
This is often used to upload signed release tarballs that can be verified back to the releasing developer/project maintainer.

It is common practice for developers in the C ecosystem to manually create additional release tarballs with slightly modified source code on their machine and upload them instead of an exact copy of the release state of the source repository.
Over time, source tarballs took the role of including build instructions for more strictly specified targets, alongside other additions such as prebuilt documentation.
For instance, the purpose of Autoconf, which was also part of the attack path for XZ Utils, is to create a configure script which is tailored specifically to the build host,
in terms of factors like the set of available machine instructions and the particularities of the running UNIX variant.
An individual building the project from source using Autoconf, will get a working binary which runs optimally on their machine.
This is the antithesis of reproducibility, because the resulting binary will of course be different depending on all the factors considered by Autoconf.
In order to prepare their software for packaging through various distributions, developers turned to publishing manually created tarballs, where build files pre-generated with Autoconf for generic target parameters are added or modified to be compatible with the broad set of intended target systems.

This process of manually creating slightly modified source code is a significant gap in reproducibility.
Of course, if we start a reproducible build from a compromised source tarball, the build process being reproducible does not help.
We need to make sure that in the scope of our work on reproducibility, we take measures that ensure build steps consume the canonical version of the source code.
A few specific recommendations for how various parties could address this issue are:
\begin{itemize}
\item GitHub and other code hosting platforms, could encourage automating the creation of modified release tarballs through CI pipelines instead of on developer machines. As a consequence, the exact contents of the tarball would be defined by auditable code in the repository.
\item GitHub and other code hosting platforms, could highlight manually uploaded artifacts as less trustworthy in their UI.
\item Developers could make the creation of modified release tarballs fully reproducible to further enhance auditability of provided tarball artifacts.
\item Developers could drop Autoconf and compilation for the native target in favor of toolchains which always cross-compile for the intended target, making modified tarballs obsolete.
\item Distributions could depend directly on source repositories instead of release tarballs.
\end{itemize}
Regarding the latter, today's VCS provide integrated measures to assure integrity of the change history as a whole, and even authenticity and non-repudiation for individual commits.
For instance, Git tracks the whole commit history in a hash chain and allows signing commits and pushes with GPG.

While distributions like Debian already mostly eliminated processes where binary software packages were built from package source tarballs and replaced them with code tracked in public VCS platforms~\cite{debian-gitpackaging}, there is still a gap between the source of packaging for a distribution and the actual upstream source code.
Our preliminary analysis of roughly 61 Debian packages, identified as the top dependencies of popular internet-facing server software on Debian, revealed that only 10 of them rely directly on the automatically generated source tarball from the release tag in the VCS.
For 9 packages, we could not identify the exact origin of the code integrated into the package.
The remaining packages rely on either extra release tarballs attached to the releases on the VCS platform or have a completely separate distribution channel for their releases.
Even some mature upstream open-source projects still do not track their code in a public VCS at all, cf.\ the statement ``I would never ever want to share any of my code before it is released'' by the maintainer of Postfix when asked about the availability of a public VCS~\cite{postfix-statement-on-vcs}.
Nevertheless, the packaging tools for Debian already support workflows for direct integration of upstream Git repositories~\cite{git-buildpackage-gitupstream}.

The same lesson does not only apply to source code.
Every link in our deployment processes and dependency chains should be accounted for with automated tools~\cite[Chapter~14]{adkins2020building}:
\begin{itemize}
\item The in-toto framework~\cite{torres2019toto} creates a cryptographically verifiable mechanism which delegates the responsibility for each link in the supply chain to a specific set of people.
It can then be used to verify that they form a chain without gaps.
\item Cloud build systems~\cite{mokhov2018build}, like Guix and Nix, could ensure that all build instructions are known and fully specified, all the way back to a bootstrapping process~\cite{bootstrappableBuildsWebsite, bitcoinBootstrap},
which for a fully reproducible dependency tree, would make that dependency tree fully verifiable via reproducibility.
\end{itemize}
Additionally, checking binary blobs into VCS, even for testing purposes, can also be viewed as a reproducibility issue,
and it would be preferable to check the commands which generate the necessary test files into VCS and execute them as part of the test setup phase.

\subsection{Code Sandboxing}
\label{sec:sandboxing}
One obvious method of impact mitigation is sandboxing of third-party library code that exhibits relevant attack surface, particularly from untrusted input.
A high-impact example of such code sandboxing is the response to the Android Stagefright vulnerability~\cite{stagefright}: media codecs were put into a tightly limited sandbox instead of being executed in-process in the relevant player/media applications~\cite{androidplatformsecuritymodel2023}.
Another example is how most modern web browsers compartmentalize their different parts, such as the network handler or renderer processes, from each other. Both Chromium~\cite{chromium-sandbox} and Firefox~\cite{firefox-sandbox} implement such sandboxes on all major operating system platforms.

All of these examples rely on custom separation strategies tweaked for their respective use-case; while the Android media sandboxing uses specific SELinux based sandbox rules and shared pages for tight compartments with zero-copy memory handling for the processed data streams but generically for different types of media codecs, browser sandboxes uses highly specialized processes for their respective use cases (e.g. using Linux seccomp-bpf and user namespace hardening~\cite{chromium-sandbox-linux}).
However, the common denominator is the basic sandbox boundary: operating system managed processes. 
Without additional hardware/OS capabilities like, e.g., CHERI~\cite{watson2015cheri}, OS processes are the most well-studied and simplest manner of sandboxing different pieces of code, and are therefore the default for sandboxing methods above the OS kernel level (explicitly not considering lower-level methods such as hardware-assisted trusted execution environments or hypervisor based approaches).

Process level sandboxing can be, and in practice often is, combined with additional compartmentalization techniques such as using separate UIDs (with Android being arguably the biggest deployment of this method~\cite{androidplatformsecuritymodel2023}), user namespaces, SELinux, seccomp, or capability dropping.

The same concept of process level sandboxing could be applied to other third-party libraries, especially including compression/decompression code. 
There are two main challenges to consider:
\begin{enumerate}
    \item \emph{Performance penalties} due to context switching, inter-pro\-cess communication, and potentially copying data buffers if zero-copy architectures cannot be used.
    This run-time overhead seems unavoidable if process boundaries are used as the basic sandboxing mechanism (as opposed to capability based proposals like CHERI that would not have such overhead, but are not currently deployed in practice), but seem practically manageable considering the global deployment within Android media handling and all modern browsers.
    \item \emph{Developer effort} is potentially much more relevant, as the previously cited sandboxes all rely on a custom design and required significant effort from the relevant projects to implement.
    This seems to be the main barrier for large-scale adoption of code sandboxing.
\end{enumerate}
One potential take-away from the xz case could therefore be to design and develop better tooling to help developers with applying process level sandboxing to their currently monolithic processes with minimal effort.
Ideally, such tooling should be embedded with the programming language, e.g., by automatically creating the relevant process forking and IPC stubs based on annotations on the level of library methods to sandbox.
To the best of our knowledge, no such generic tools exist at the moment, although some proposals, e.g., for accessing unsafe native code from Rust~\cite{10.1145/3144555.3144562} or for JavaScript native code libraries~\cite{10.1145/3607199.3607233}, exist. 
The Android ``isolated process''~\cite{android-isolated-process} might come closest to a generic mechanism that is already broadly available, but, while it offers a generic UID-separated process sandbox, the relevant IPC with sandboxed processes based on AIDL still needs to be adapted for each use case.
Similar developer effort is required to use other open source tools like Sandbox2~\cite{sandbox2}.

Summarizing the state-of-the-art in library sandboxing, the basic techniques for tight compartmentalization of potentially untrusted third-party code are well established with manageable performance overhead and mostly based on process level separation.
However, for adoption beyond obvious high-impact use cases, better developer tooling will be required to minimize developer effort in applying such sandboxing.

\subsection{Legal Defenses}
\label{sec:legaldefenses}
Another potential mitigation that, to the best of our knowledge, has so far not been utilized is the established legal system. 
Under the assumption that the attacker identity (or, presumably, identities of a group of attackers acting together) can be established in \emph{post-factum} investigation, a court of law could establish if any actions involved in such an attack were committed with the intent to cause harm. 
That is, even if an attack is ultimately unsuccessful or the actual exploit is never executed,\footnote{As in this particular case, where we currently have no evidence that the specific OpenSSH exploit was actually triggered by the attacker(s).} putting critical infrastructure components at significant risk could be found to be criminal behavior.\footnote{This is not and should not be understood as legal analysis, but we refer to established international statutes like Art.~8, §2(b)(ii) of the Rome Statute of the International Criminal Court~\cite{rome-statute}: ``Intentionally directing attacks against civilian objects, that is, objects which are not military objectives.''}
While we are well aware that such law is typically not applied to intelligence service or other national security actors, we intentionally pose the question if it should be used in future cases as a form of deterrence against the more egregious attacks on software, hardware, and services in common use that have significant risk of collateral damage for large parts of the population.
As with other legal defenses, this would not be a mechanism of prevention, but risk of punishment could still act as a further mitigation.

\section{Conclusion}
\label{sec:conclusion}
The XZ Utils backdoor is another example of open source software that has become a central component of (too?) many foundational services, and therefore of critical Internet infrastructure components. 
Through the complexity and duration of the full attack path, we can speculate that the threat actor(s) are part of a highly sophisticated organization with stable funding. 
This example can be used as an excellent teaching opportunity, both in terms of social/organizational and technical aspects. 
We have very briefly summarized what we consider to be the critical components of the attack path, referencing many more detailed resources for specific parts of the attack analysis. 

From this attack path, we speculate about potential mitigations that could have broken this chain. 
However, while we try to derive generic potential improvements for open source projects to mitigate the different parts that this particular attack example relied on, we note that other attack paths will most probably still exist even if all our suggested mitigations are implemented in the future. 
Particularly the complexity of old build toolchains around programming languages without memory safety will continue to leave open many different attack paths~\cite{project-zero-linux-bug} that are hard to find, difficult to audit, and provide the cover of plausible deniability for attackers in the form of exploitable ``bugdoor'' attacks\footnote{The portmanteau ``bugdoor'' describes a backdoor---typically implanted in the source code---that can plausibly be argued to have been a genuine bug in the relevant code. It is unclear when and by whom this term was coined, but early sources include an analysis on embedded systems firmware from 2014~\cite{10.1145/2664243.2664268}. The concept itself is generic and not limited to pure memory safety issues, as formalized (but not explicitly named as potential bugdoor vectors) in the seminal ``Weird Machines'' work~\cite{8226852}.}. 
Moving to memory safe languages, particularly with declarative build system specifications instead of arbitrary interpreted scripts executed during the build process, has many advantages for general code quality besides the hope of making these kinds of attacks somewhat harder (but not impossible).

One particular take-away is a strengthened belief voiced by many of the experts involved with analyzing this case that similar attacks are possible on proprietary, closed source codebases, and that, as a security community, we have to assume other attacks with comparable complexity and sophistication to be actively deployed in multiple projects/products.
Open source makes it possible to focus a diverse set of experts on an investigation and is therefore considered a major strength for quick analysis.

\begin{acks}
This work has been carried out within the scope of Digidow, the Christian Doppler Laboratory for Private Digital Authentication in the Physical World and has partially been supported by the LIT Secure and Correct Systems Lab.
We gratefully acknowledge financial support by the Austrian Federal Ministry of Labour and Economy, the National Foundation for Research, Technology and Development, the Christian Doppler Research Association, 3 Banken IT GmbH, ekey biometric systems GmbH, Kepler Universit\"atsklinikum GmbH, NXP Semiconductors Austria GmbH \& Co KG, \"Osterreichische Staatsdruckerei GmbH, and the State of Upper Austria.
\end{acks}

\bibliographystyle{ACM-Reference-Format}
\bibliography{main}


\begin{thebibliography}{56}


\ifx \showCODEN    \undefined \def \showCODEN     #1{\unskip}     \fi
\ifx \showDOI      \undefined \def \showDOI       #1{#1}\fi
\ifx \showISBNx    \undefined \def \showISBNx     #1{\unskip}     \fi
\ifx \showISBNxiii \undefined \def \showISBNxiii  #1{\unskip}     \fi
\ifx \showISSN     \undefined \def \showISSN      #1{\unskip}     \fi
\ifx \showLCCN     \undefined \def \showLCCN      #1{\unskip}     \fi
\ifx \shownote     \undefined \def \shownote      #1{#1}          \fi
\ifx \showarticletitle \undefined \def \showarticletitle #1{#1}   \fi
\ifx \showURL      \undefined \def \showURL       {\relax}        \fi
\providecommand\bibfield[2]{#2}
\providecommand\bibinfo[2]{#2}
\providecommand\natexlab[1]{#1}
\providecommand\showeprint[2][]{arXiv:#2}

\bibitem[Abbadini et~al\mbox{.}(2023)]%
        {10.1145/3607199.3607233}
\bibfield{author}{\bibinfo{person}{Marco Abbadini}, \bibinfo{person}{Dario
  Facchinetti}, \bibinfo{person}{Gianluca Oldani}, \bibinfo{person}{Matthew
  Rossi}, {and} \bibinfo{person}{Stefano Paraboschi}.}
  \bibinfo{year}{2023}\natexlab{}.
\newblock \showarticletitle{{NatiSand: Native Code Sandboxing for JavaScript
  Runtimes}}. In \bibinfo{booktitle}{\emph{Proceedings of the 26th
  International Symposium on Research in Attacks, Intrusions and Defenses}}
  (Hong Kong, China) \emph{(\bibinfo{series}{RAID '23})}.
  \bibinfo{publisher}{ACM}, \bibinfo{pages}{639--653}.
\newblock
\urldef\tempurl%
\url{https://doi.org/10.1145/3607199.3607233}
\showDOI{\tempurl}


\bibitem[Adkins et~al\mbox{.}(2020)]%
        {adkins2020building}
\bibfield{author}{\bibinfo{person}{Heather Adkins}, \bibinfo{person}{Betsy
  Beyer}, \bibinfo{person}{Paul Blankinship}, \bibinfo{person}{Piotr
  Lewandowski}, \bibinfo{person}{Ana Oprea}, {and} \bibinfo{person}{Adam
  Stubblefield}.} \bibinfo{year}{2020}\natexlab{}.
\newblock \bibinfo{booktitle}{\emph{{Building Secure and Reliable Systems}}}.
\newblock \bibinfo{publisher}{O'Reilly Media}.
\newblock
\showISBNx{978-1492083122}
\urldef\tempurl%
\url{https://google.github.io/building-secure-and-reliable-systems/}
\showURL{%
\tempurl}


\bibitem[{Akamai Security Intelligence Group}(2024)]%
        {akamai2024}
\bibfield{author}{\bibinfo{person}{{Akamai Security Intelligence Group}}.}
  \bibinfo{year}{2024}\natexlab{}.
\newblock \bibinfo{title}{{XZ Utils Backdoor — Everything You Need to Know,
  and What You Can Do}}.
\newblock
\newblock
\urldef\tempurl%
\url{https://www.akamai.com/blog/security-research/critical-linux-backdoor-xz-utils-discovered-what-to-know}
\showURL{%
\tempurl}
\newblock
\shownote{(accessed: 2024-04-09)}.


\bibitem[{AOSP}(2024)]%
        {android-isolated-process}
\bibfield{author}{\bibinfo{person}{{AOSP}}.} \bibinfo{year}{2024}\natexlab{}.
\newblock \bibinfo{title}{{android:isolatedProcess feature}}.
\newblock
\newblock
\urldef\tempurl%
\url{https://developer.android.com/guide/topics/manifest/service-element#isolated}
\showURL{%
\tempurl}
\newblock
\shownote{(accessed: 2024-04-08)}.


\bibitem[Boehs(2024)]%
        {boehs2024}
\bibfield{author}{\bibinfo{person}{Evan Boehs}.}
  \bibinfo{year}{2024}\natexlab{}.
\newblock \bibinfo{title}{{Everything I know about the XZ backdoor}}.
\newblock
\newblock
\urldef\tempurl%
\url{https://boehs.org/node/everything-i-know-about-the-xz-backdoor}
\showURL{%
\tempurl}
\newblock
\shownote{(accessed: 2024-04-02)}.


\bibitem[{Bootstrappable Builds}(2017)]%
        {bootstrappableBuildsWebsite}
\bibfield{author}{\bibinfo{person}{{Bootstrappable Builds}}.}
  \bibinfo{year}{2017}\natexlab{}.
\newblock \bibinfo{title}{Bootstrappable Builds Website}.
\newblock
\newblock
\urldef\tempurl%
\url{https://bootstrappable.org/}
\showURL{%
\tempurl}
\newblock
\shownote{(accessed: 2024-04-04)}.


\bibitem[Chacon and Straub(2024)]%
        {git-signing2024}
\bibfield{author}{\bibinfo{person}{Scott Chacon} {and} \bibinfo{person}{Ben
  Straub}.} \bibinfo{year}{2024}\natexlab{}.
\newblock \showarticletitle{{Git Tools - Signing Your Work}}.
\newblock In \bibinfo{booktitle}{\emph{Pro Git} (\bibinfo{edition}{2nd} ed.)}.
  \bibinfo{publisher}{Apress}.
\newblock
\urldef\tempurl%
\url{https://git-scm.com/book/en/v2/Git-Tools-Signing-Your-Work}
\showURL{%
\tempurl}
\newblock
\shownote{(accessed: 2024-04-12)}.


\bibitem[Coldwind(2024)]%
        {coldwind2024}
\bibfield{author}{\bibinfo{person}{Gynvael Coldwind}.}
  \bibinfo{year}{2024}\natexlab{}.
\newblock \bibinfo{title}{{xz/liblzma: Bash-stage Obfuscation Explained}}.
\newblock
\newblock
\urldef\tempurl%
\url{https://gynvael.coldwind.pl/?lang=en&id=782}
\showURL{%
\tempurl}
\newblock
\shownote{(accessed: 2024-04-02)}.


\bibitem[Collin(2022)]%
        {mailarchive22022}
\bibfield{author}{\bibinfo{person}{Lasse Collin}.}
  \bibinfo{year}{2022}\natexlab{}.
\newblock \bibinfo{title}{{Re: [xz-devel] XZ for Java}}.
\newblock \bibinfo{howpublished}{Reply on mailing list xz-devel}.
\newblock
\urldef\tempurl%
\url{https://www.mail-archive.com/xz-devel@tukaani.org/msg00567.html}
\showURL{%
\tempurl}
\newblock
\shownote{(accessed: 2024-04-11)}.


\bibitem[Collin(2024)]%
        {collin2024}
\bibfield{author}{\bibinfo{person}{Lasse Collin}.}
  \bibinfo{year}{2024}\natexlab{}.
\newblock \bibinfo{title}{{XZ Utils backdoor}}.
\newblock
\newblock
\urldef\tempurl%
\url{https://tukaani.org/xz-backdoor/}
\showURL{%
\tempurl}
\newblock
\shownote{(accessed: 2024-04-12)}.


\bibitem[Cox(2024)]%
        {cox2024}
\bibfield{author}{\bibinfo{person}{Russ Cox}.} \bibinfo{year}{2024}\natexlab{}.
\newblock \bibinfo{title}{{The xz attack shell script}}.
\newblock
\newblock
\urldef\tempurl%
\url{https://research.swtch.com/xz-script}
\showURL{%
\tempurl}
\newblock
\shownote{(accessed: 2024-04-11)}.


\bibitem[Dong(2019)]%
        {bitcoinBootstrap}
\bibfield{author}{\bibinfo{person}{Carl Dong}.}
  \bibinfo{year}{2019}\natexlab{}.
\newblock \bibinfo{title}{Bitcoin Build System Security}.
\newblock \bibinfo{howpublished}{Talk at Breaking Bitcoin 2019 Amsterdam}.
\newblock
\urldef\tempurl%
\url{https://www.youtube.com/watch?v=I2iShmUTEl8}
\showURL{%
\tempurl}
\newblock
\shownote{(accessed 2024-04-12)}.


\bibitem[Drake(2016)]%
        {stagefright}
\bibfield{author}{\bibinfo{person}{Joshua~J. Drake}.}
  \bibinfo{year}{2016}\natexlab{}.
\newblock \showarticletitle{Stagefright: An Android Exploitation Case Study}.
  In \bibinfo{booktitle}{\emph{{WOOT'16: 10th USENIX Workshop on Offensive
  Technologies}}} (Austin, TX). \bibinfo{publisher}{USENIX Association}.
\newblock
\urldef\tempurl%
\url{https://www.usenix.org/conference/woot16/workshop-program/presentation/drake}
\showURL{%
\tempurl}


\bibitem[Dullien(2020)]%
        {8226852}
\bibfield{author}{\bibinfo{person}{Thomas Dullien}.}
  \bibinfo{year}{2020}\natexlab{}.
\newblock \showarticletitle{{Weird Machines, Exploitability, and Provable
  Unexploitability}}.
\newblock \bibinfo{journal}{\emph{IEEE Transactions on Emerging Topics in
  Computing}} \bibinfo{volume}{8}, \bibinfo{number}{2} (\bibinfo{year}{2020}),
  \bibinfo{pages}{391--403}.
\newblock
\urldef\tempurl%
\url{https://doi.org/10.1109/TETC.2017.2785299}
\showDOI{\tempurl}


\bibitem[{European Commission}(2024)]%
        {eidas2024}
\bibfield{author}{\bibinfo{person}{{European Commission}}.}
  \bibinfo{year}{2024}\natexlab{}.
\newblock \bibinfo{title}{{eIDAS Regulation}}.
\newblock
\newblock
\urldef\tempurl%
\url{https://digital-strategy.ec.europa.eu/en/policies/eidas-regulation}
\showURL{%
\tempurl}
\newblock
\shownote{(accessed: 2024-04-12)}.


\bibitem[Freund(2024)]%
        {freund2024}
\bibfield{author}{\bibinfo{person}{Andres Freund}.}
  \bibinfo{year}{2024}\natexlab{}.
\newblock \bibinfo{title}{{backdoor in upstream xz/liblzma leading to ssh
  server compromise}}.
\newblock \bibinfo{howpublished}{Post on mailing list oss-security@openwall}.
\newblock
\urldef\tempurl%
\url{https://openwall.com/lists/oss-security/2024/03/29/4}
\showURL{%
\tempurl}
\newblock
\shownote{(accessed: 2024-04-12)}.


\bibitem[Gabriel et~al\mbox{.}(2021)]%
        {debian-gitpackaging}
\bibfield{author}{\bibinfo{person}{Mike Gabriel}, \bibinfo{person}{Carsten
  Schoenert}, \bibinfo{person}{Raphael Hertzog}, {and} \bibinfo{person}{Brian
  Thompson}.} \bibinfo{year}{2021}\natexlab{}.
\newblock \bibinfo{title}{{GitPackaging}}.
\newblock \bibinfo{howpublished}{Debian Wiki}.
\newblock
\urldef\tempurl%
\url{https://wiki.debian.org/GitPackaging}
\showURL{%
\tempurl}
\newblock
\shownote{(accessed: 2024-04-10)}.


\bibitem[{GitHub, Inc.}(2024)]%
        {githubbranchprotection2024}
\bibfield{author}{\bibinfo{person}{{GitHub, Inc.}}}
  \bibinfo{year}{2024}\natexlab{}.
\newblock \bibinfo{title}{{About protected branches}}.
\newblock
\newblock
\urldef\tempurl%
\url{https://docs.github.com/en/repositories/configuring-branches-and-merges-in-your-repository/managing-protected-branches/about-protected-branches}
\showURL{%
\tempurl}
\newblock
\shownote{(accessed: 2024-04-12)}.


\bibitem[{Google}(2013)]%
        {ossfuzz2024}
\bibfield{author}{\bibinfo{person}{{Google}}.} \bibinfo{year}{2013}\natexlab{}.
\newblock \bibinfo{title}{{OSS-Fuzz: Continuous Fuzzing for Open Source
  Software}}.
\newblock
\newblock
\urldef\tempurl%
\url{https://github.com/google/oss-fuzz}
\showURL{%
\tempurl}
\newblock
\shownote{(accessed: 2024-04-11)}.


\bibitem[{Google}(2024a)]%
        {ossfuzznewproject2024}
\bibfield{author}{\bibinfo{person}{{Google}}.}
  \bibinfo{year}{2024}\natexlab{a}.
\newblock \bibinfo{title}{{Accepting New Projects}}.
\newblock
\newblock
\urldef\tempurl%
\url{https://google.github.io/oss-fuzz/getting-started/accepting-new-projects/}
\showURL{%
\tempurl}
\newblock
\shownote{(accessed: 2024-04-12)}.


\bibitem[{Google}(2024b)]%
        {sandbox2}
\bibfield{author}{\bibinfo{person}{{Google}}.}
  \bibinfo{year}{2024}\natexlab{b}.
\newblock \bibinfo{title}{Sandbox2 Explained}.
\newblock
\newblock
\urldef\tempurl%
\url{https://developers.google.com/code-sandboxing/sandbox2/explained}
\showURL{%
\tempurl}


\bibitem[{Google Chromium Team}(2024a)]%
        {chromium-sandbox-linux}
\bibfield{author}{\bibinfo{person}{{Google Chromium Team}}.}
  \bibinfo{year}{2024}\natexlab{a}.
\newblock \bibinfo{title}{{Linux Sandboxing}}.
\newblock
\newblock
\urldef\tempurl%
\url{https://chromium.googlesource.com/chromium/src/+/0e94f26e8/docs/linux_sandboxing.md}
\showURL{%
\tempurl}


\bibitem[{Google Chromium Team}(2024b)]%
        {chromium-sandbox}
\bibfield{author}{\bibinfo{person}{{Google Chromium Team}}.}
  \bibinfo{year}{2024}\natexlab{b}.
\newblock \bibinfo{title}{{Sandbox}}.
\newblock
\newblock
\urldef\tempurl%
\url{https://chromium.googlesource.com/chromium/src/+/b4730a0c2773d8f6728946013eb812c6d3975bec/docs/design/sandbox.md}
\showURL{%
\tempurl}


\bibitem[Günther(2021)]%
        {git-buildpackage-gitupstream}
\bibfield{author}{\bibinfo{person}{Guido Günther}.}
  \bibinfo{year}{2021}\natexlab{}.
\newblock \bibinfo{title}{{When upstream uses Git}}.
\newblock \bibinfo{howpublished}{git-buildpackage Documentation Version
  0.9.31}.
\newblock
\urldef\tempurl%
\url{https://honk.sigxcpu.org/projects/git-buildpackage/manual-html/gbp.import.upstream-git.html}
\showURL{%
\tempurl}
\newblock
\shownote{(accessed: 2024-04-10)}.


\bibitem[Horn(2021)]%
        {project-zero-linux-bug}
\bibfield{author}{\bibinfo{person}{Jann Horn}.}
  \bibinfo{year}{2021}\natexlab{}.
\newblock \bibinfo{title}{{How a simple Linux kernel memory corruption bug can
  lead to complete system compromise}}.
\newblock \bibinfo{howpublished}{Project Zero}.
\newblock
\urldef\tempurl%
\url{https://googleprojectzero.blogspot.com/2021/10/how-simple-linux-kernel-memory.html}
\showURL{%
\tempurl}
\newblock
\shownote{(accessed: 2024-04-12)}.


\bibitem[{International Criminal Court}(2021)]%
        {rome-statute}
\bibfield{author}{\bibinfo{person}{{International Criminal Court}}.}
  \bibinfo{year}{2021}\natexlab{}.
\newblock \bibinfo{booktitle}{\emph{{Rome Statute of the International Criminal
  Court}}}.
\newblock \bibinfo{publisher}{International Criminal Court},
  \bibinfo{address}{The Hague, The Netherlands}.
\newblock
\showISBNx{92-9227-386-8}
\urldef\tempurl%
\url{https://www.icc-cpi.int/sites/default/files/Publications/Rome-Statute.pdf}
\showURL{%
\tempurl}


\bibitem[Jansen(2023a)]%
        {ifuncpatch22023}
\bibfield{author}{\bibinfo{person}{Hans Jansen}.}
  \bibinfo{year}{2023}\natexlab{a}.
\newblock \bibinfo{title}{{Add ifunc check to CMakeLists.txt}}.
\newblock
\newblock
\urldef\tempurl%
\url{https://git.tukaani.org/?p=xz.git;a=commitdiff;h=b72d21202402a603db6d512fb9271cfa83249639}
\showURL{%
\tempurl}
\newblock
\shownote{(accessed: 2024-04-12)}.


\bibitem[Jansen(2023b)]%
        {ifuncpatch12023}
\bibfield{author}{\bibinfo{person}{Hans Jansen}.}
  \bibinfo{year}{2023}\natexlab{b}.
\newblock \bibinfo{title}{{Add ifunc check to configure.ac}}.
\newblock
\newblock
\urldef\tempurl%
\url{https://git.tukaani.org/?p=xz.git;a=commitdiff;h=23b5c36fb71904bfbe16bb20f976da38dadf6c3b}
\showURL{%
\tempurl}
\newblock
\shownote{(accessed: 2024-04-12)}.


\bibitem[Jansen(2024)]%
        {debianbug2024}
\bibfield{author}{\bibinfo{person}{Hans Jansen}.}
  \bibinfo{year}{2024}\natexlab{}.
\newblock \bibinfo{title}{{xz-utils: New upstream version available}}.
\newblock \bibinfo{howpublished}{Debian Bug report \#1067708}.
\newblock
\urldef\tempurl%
\url{https://bugs.debian.org/cgi-bin/bugreport.cgi?bug=1067708}
\showURL{%
\tempurl}
\newblock
\shownote{(accessed: 2024-04-11)}.


\bibitem[{JiaT75}(2021)]%
        {prlibarchive2021}
\bibfield{author}{\bibinfo{person}{{JiaT75}}.} \bibinfo{year}{2021}\natexlab{}.
\newblock \bibinfo{title}{{Added error text to warning when untaring with
  bsdtar}}.
\newblock \bibinfo{howpublished}{Pull request \#1609}.
\newblock
\urldef\tempurl%
\url{https://github.com/libarchive/libarchive/pull/1609}
\showURL{%
\tempurl}
\newblock
\shownote{(accessed: 2024-04-02)}.


\bibitem[{JiaT75}(2022)]%
        {firstcommitasmaintainer2022}
\bibfield{author}{\bibinfo{person}{{JiaT75}}.} \bibinfo{year}{2022}\natexlab{}.
\newblock \bibinfo{title}{{CMake: Update .gitignore for CMake artifacts from in
  source build}}.
\newblock
\newblock
\urldef\tempurl%
\url{https://github.com/tukaani-project/xz/commit/8ace358d65059152d9a1f43f4770170d29d35754}
\showURL{%
\tempurl}
\newblock
\shownote{(accessed: 2024-04-11)}.


\bibitem[{JiaT75}(2023a)]%
        {oss-fuzz-pull-request-disable-ifunc}
\bibfield{author}{\bibinfo{person}{{JiaT75}}.}
  \bibinfo{year}{2023}\natexlab{a}.
\newblock \bibinfo{title}{{xz: Disable ifunc to fix Issue 60259}}.
\newblock \bibinfo{howpublished}{Pull request \#10667}.
\newblock
\urldef\tempurl%
\url{https://github.com/google/oss-fuzz/pull/10667}
\showURL{%
\tempurl}
\newblock
\shownote{(accessed: 2024-04-12)}.


\bibitem[{JiaT75}(2023b)]%
        {oss-fuzz-pull-request-add-jiat75-to-contacts}
\bibfield{author}{\bibinfo{person}{{JiaT75}}.}
  \bibinfo{year}{2023}\natexlab{b}.
\newblock \bibinfo{title}{{xz-java: Add upstream maintainers to contact fields
  in project.yaml}}.
\newblock \bibinfo{howpublished}{Pull request \#11295}.
\newblock
\urldef\tempurl%
\url{https://github.com/google/oss-fuzz/pull/11295/}
\showURL{%
\tempurl}
\newblock
\shownote{(accessed: 2024-04-12)}.


\bibitem[{JiaT75}(2023c)]%
        {prprimarycontact2023}
\bibfield{author}{\bibinfo{person}{{JiaT75}}.}
  \bibinfo{year}{2023}\natexlab{c}.
\newblock \bibinfo{title}{{XZ updates}}.
\newblock \bibinfo{howpublished}{Pull request \#9960}.
\newblock
\urldef\tempurl%
\url{https://github.com/google/oss-fuzz/pull/9960}
\showURL{%
\tempurl}
\newblock
\shownote{(accessed: 2024-04-11)}.


\bibitem[{jonathanmetzman}(2024)]%
        {oss-fuzz-pull-request-disable-ifunc-comment}
\bibfield{author}{\bibinfo{person}{{jonathanmetzman}}.}
  \bibinfo{year}{2024}\natexlab{}.
\newblock \bibinfo{title}{{xz: Disable ifunc to fix Issue 60259}}.
\newblock \bibinfo{howpublished}{Comment on pull request \#10667}.
\newblock
\urldef\tempurl%
\url{https://github.com/google/oss-fuzz/pull/10667\#pullrequestreview-1518981986}
\showURL{%
\tempurl}
\newblock
\shownote{(accessed: 2024-04-12)}.


\bibitem[Kumar(2022)]%
        {mailarchive32022}
\bibfield{author}{\bibinfo{person}{Jigar Kumar}.}
  \bibinfo{year}{2022}\natexlab{}.
\newblock \bibinfo{title}{{Re: [xz-devel] [PATCH] String to filter and filter
  to string}}.
\newblock \bibinfo{howpublished}{Reply on mailing list xz-devel}.
\newblock
\urldef\tempurl%
\url{https://www.mail-archive.com/xz-devel@tukaani.org/msg00555.html}
\showURL{%
\tempurl}
\newblock
\shownote{(accessed: 2024-04-12)}.


\bibitem[Lamowski et~al\mbox{.}(2017)]%
        {10.1145/3144555.3144562}
\bibfield{author}{\bibinfo{person}{Benjamin Lamowski}, \bibinfo{person}{Carsten
  Weinhold}, \bibinfo{person}{Adam Lackorzynski}, {and}
  \bibinfo{person}{Hermann H\"{a}rtig}.} \bibinfo{year}{2017}\natexlab{}.
\newblock \showarticletitle{{Sandcrust: Automatic Sandboxing of Unsafe
  Components in Rust}}. In \bibinfo{booktitle}{\emph{Proceedings of the 9th
  Workshop on Programming Languages and Operating Systems}} (Shanghai, China)
  \emph{(\bibinfo{series}{PLOS '17})}. \bibinfo{publisher}{ACM},
  \bibinfo{pages}{51--57}.
\newblock
\urldef\tempurl%
\url{https://doi.org/10.1145/3144555.3144562}
\showDOI{\tempurl}


\bibitem[Lins et~al\mbox{.}(2023)]%
        {2023-lins-nordsec}
\bibfield{author}{\bibinfo{person}{Mario Lins}, \bibinfo{person}{René
  Mayrhofer}, \bibinfo{person}{Michael Roland}, {and}
  \bibinfo{person}{Alastair~R. Beresford}.} \bibinfo{year}{2023}\natexlab{}.
\newblock \showarticletitle{{Mobile App Distribution Transparency (MADT):
  Design and evaluation of a system to mitigate necessary trust in mobile app
  distribution systems}}. In \bibinfo{booktitle}{\emph{Secure IT Systems. 28th
  Nordic Conference, NordSec 2023}} (Oslo, Norway)
  \emph{(\bibinfo{series}{LNCS}, Vol.~\bibinfo{volume}{14324/2024})}.
  \bibinfo{publisher}{Springer}, \bibinfo{pages}{185--203}.
\newblock
\urldef\tempurl%
\url{https://doi.org/10.1007/978-3-031-47748-5\_11}
\showDOI{\tempurl}


\bibitem[Mayrhofer et~al\mbox{.}(2024)]%
        {androidplatformsecuritymodel2023}
\bibfield{author}{\bibinfo{person}{René Mayrhofer},
  \bibinfo{person}{Jeffrey~Vander Stoep}, \bibinfo{person}{Chad Brubaker},
  \bibinfo{person}{Dianne Hackborn}, \bibinfo{person}{Bram Bonné},
  \bibinfo{person}{Güliz~Seray Tuncay}, \bibinfo{person}{Roger~Piqueras
  Jover}, {and} \bibinfo{person}{Michael~A. Specter}.}
  \bibinfo{year}{2024}\natexlab{}.
\newblock \bibinfo{booktitle}{\emph{{The Android Platform Security Model
  (2023)}}}.
\newblock
\urldef\tempurl%
\url{https://doi.org/10.48550/arXiv.1904.05572}
\showDOI{\tempurl}


\bibitem[Midha and Bhattacherjee(2012)]%
        {MIDHA201223}
\bibfield{author}{\bibinfo{person}{Vishal Midha} {and} \bibinfo{person}{Anol
  Bhattacherjee}.} \bibinfo{year}{2012}\natexlab{}.
\newblock \showarticletitle{{Governance practices and software maintenance: A
  study of open source projects}}.
\newblock \bibinfo{journal}{\emph{Decision Support Systems}}
  \bibinfo{volume}{54}, \bibinfo{number}{1} (\bibinfo{year}{2012}),
  \bibinfo{pages}{23--32}.
\newblock
\urldef\tempurl%
\url{https://doi.org/10.1016/j.dss.2012.03.002}
\showDOI{\tempurl}


\bibitem[Mokhov et~al\mbox{.}(2018)]%
        {mokhov2018build}
\bibfield{author}{\bibinfo{person}{Andrey Mokhov}, \bibinfo{person}{Neil
  Mitchell}, {and} \bibinfo{person}{Simon Peyton~Jones}.}
  \bibinfo{year}{2018}\natexlab{}.
\newblock \showarticletitle{Build Systems à la Carte}.
\newblock \bibinfo{journal}{\emph{Proc. ACM Program.}} \bibinfo{volume}{2},
  \bibinfo{number}{ICFP}, Article \bibinfo{articleno}{79}
  (\bibinfo{year}{2018}), \bibinfo{numpages}{29}~pages.
\newblock
\urldef\tempurl%
\url{https://doi.org/10.1145/3236774}
\showDOI{\tempurl}


\bibitem[{Mozilla}(2024)]%
        {cargovet2024}
\bibfield{author}{\bibinfo{person}{{Mozilla}}.}
  \bibinfo{year}{2024}\natexlab{}.
\newblock \bibinfo{title}{{Cargo Vet}}.
\newblock
\newblock
\urldef\tempurl%
\url{https://mozilla.github.io/cargo-vet/}
\showURL{%
\tempurl}
\newblock
\shownote{(accessed: 2024-04-12)}.


\bibitem[{Mozilla Wiki}(2024)]%
        {firefox-sandbox}
\bibfield{author}{\bibinfo{person}{{Mozilla Wiki}}.}
  \bibinfo{year}{2024}\natexlab{}.
\newblock \bibinfo{title}{{Security/Sandbox/Process model}}.
\newblock
\newblock
\urldef\tempurl%
\url{https://wiki.mozilla.org/Security/Sandbox/Process_model}
\showURL{%
\tempurl}


\bibitem[O'Donell(2024)]%
        {ifunc2024}
\bibfield{author}{\bibinfo{person}{Carlos O'Donell}.}
  \bibinfo{year}{2024}\natexlab{}.
\newblock \bibinfo{title}{{GNU\_IFUNC}}.
\newblock \bibinfo{howpublished}{glibc wiki}.
\newblock
\urldef\tempurl%
\url{https://sourceware.org/glibc/wiki/GNU_IFUNC}
\showURL{%
\tempurl}
\newblock
\shownote{(accessed: 2024-04-10)}.


\bibitem[{Open Worldwide Application Security Project (OWASP)}(2023)]%
        {owasp-dev-guide-oss}
\bibfield{author}{\bibinfo{person}{{Open Worldwide Application Security Project
  (OWASP)}}.} \bibinfo{year}{2023}\natexlab{}.
\newblock \bibinfo{booktitle}{\emph{{OWASP Developer Guide}}}.
\newblock \bibinfo{type}{Release version v4.0.1}.
\newblock
\urldef\tempurl%
\url{https://owasp.org/www-project-developer-guide/release/verification/dos_donts/open_source_software/}
\showURL{%
\tempurl}
\newblock
\shownote{(accessed: 2024-04-12)}.


\bibitem[{Reproducible Builds}(2016)]%
        {reproducibleBuildsWebsite}
\bibfield{author}{\bibinfo{person}{{Reproducible Builds}}.}
  \bibinfo{year}{2016}\natexlab{}.
\newblock \bibinfo{title}{Reproducible Builds Website}.
\newblock
\newblock
\urldef\tempurl%
\url{https://reproducible-builds.org/}
\showURL{%
\tempurl}
\newblock
\shownote{(accessed: 2024-04-12)}.


\bibitem[Souppaya et~al\mbox{.}(2022)]%
        {nist-sp800-218}
\bibfield{author}{\bibinfo{person}{Murugiah Souppaya}, \bibinfo{person}{Karen
  Scarfone}, {and} \bibinfo{person}{Donna Dodson}.}
  \bibinfo{year}{2022}\natexlab{}.
\newblock \bibinfo{booktitle}{\emph{{Secure Software Development Framework
  (SSDF) Version 1.1: Recommendations for Mitigating the Risk of Software
  Vulnerabilities}}}.
\newblock \bibinfo{type}{NIST Special Publication 800-218}.
  \bibinfo{institution}{NIST}.
\newblock
\urldef\tempurl%
\url{https://doi.org/10.6028/NIST.SP.800-218}
\showDOI{\tempurl}


\bibitem[Tan(2022)]%
        {mailarchive2022}
\bibfield{author}{\bibinfo{person}{Jia Tan}.} \bibinfo{year}{2022}\natexlab{}.
\newblock \bibinfo{title}{{[xz-devel] [PATCH] String to filter and filter to
  string}}.
\newblock \bibinfo{howpublished}{Post on mailing list xz-devel}.
\newblock
\urldef\tempurl%
\url{https://www.mail-archive.com/xz-devel@tukaani.org/msg00553.html}
\showURL{%
\tempurl}
\newblock
\shownote{(accessed: 2024-04-11)}.


\bibitem[Tan(2024a)]%
        {backdoor2024}
\bibfield{author}{\bibinfo{person}{Jia Tan}.} \bibinfo{year}{2024}\natexlab{a}.
\newblock \bibinfo{title}{{Tests: Add a few test files}}.
\newblock
\newblock
\urldef\tempurl%
\url{https://git.tukaani.org/?p=xz.git;a=commitdiff;h=cf44e4b7f5dfdbf8c78aef377c10f71e274f63c0}
\showURL{%
\tempurl}
\newblock
\shownote{(accessed: 2024-04-11)}.


\bibitem[Tan(2024b)]%
        {updatebackdoor2024}
\bibfield{author}{\bibinfo{person}{Jia Tan}.} \bibinfo{year}{2024}\natexlab{b}.
\newblock \bibinfo{title}{{Tests: Update two test files}}.
\newblock
\newblock
\urldef\tempurl%
\url{https://git.tukaani.org/?p=xz.git;a=commitdiff;h=6e636819e8f070330d835fce46289a3ff72a7b89}
\showURL{%
\tempurl}
\newblock
\shownote{(accessed: 2024-04-11)}.


\bibitem[Tan et~al\mbox{.}(2014)]%
        {10.1145/2664243.2664268}
\bibfield{author}{\bibinfo{person}{Samuel~Junjie Tan}, \bibinfo{person}{Sergey
  Bratus}, {and} \bibinfo{person}{Travis Goodspeed}.}
  \bibinfo{year}{2014}\natexlab{}.
\newblock \showarticletitle{{Interrupt-oriented bugdoor programming: a
  minimalist approach to bugdooring embedded systems firmware}}. In
  \bibinfo{booktitle}{\emph{Proceedings of the 30th Annual Computer Security
  Applications Conference}} (New Orleans, Louisiana, USA)
  \emph{(\bibinfo{series}{ACSAC '14})}. \bibinfo{publisher}{ACM},
  \bibinfo{pages}{116--125}.
\newblock
\urldef\tempurl%
\url{https://doi.org/10.1145/2664243.2664268}
\showDOI{\tempurl}


\bibitem[{The Tukaani Project}(2024)]%
        {lzma2024}
\bibfield{author}{\bibinfo{person}{{The Tukaani Project}}.}
  \bibinfo{year}{2024}\natexlab{}.
\newblock \bibinfo{title}{{LZMA Utils}}.
\newblock
\newblock
\urldef\tempurl%
\url{https://tukaani.org/lzma/}
\showURL{%
\tempurl}
\newblock
\shownote{(accessed: 2024-04-02)}.


\bibitem[Torres-Arias et~al\mbox{.}(2019)]%
        {torres2019toto}
\bibfield{author}{\bibinfo{person}{Santiago Torres-Arias},
  \bibinfo{person}{Hammad Afzali}, \bibinfo{person}{Trishank~Karthik
  Kuppusamy}, \bibinfo{person}{Reza Curtmola}, {and} \bibinfo{person}{Justin
  Cappos}.} \bibinfo{year}{2019}\natexlab{}.
\newblock \showarticletitle{in-toto: Providing farm-to-table guarantees for
  bits and bytes}. In \bibinfo{booktitle}{\emph{28th USENIX Security Symposium
  (USENIX Security 19)}} (Santa Clara, CA). \bibinfo{publisher}{USENIX
  Association}, \bibinfo{pages}{1393--1410}.
\newblock
\urldef\tempurl%
\url{https://www.usenix.org/conference/usenixsecurity19/presentation/torres-arias}
\showURL{%
\tempurl}


\bibitem[Venema(2014)]%
        {postfix-statement-on-vcs}
\bibfield{author}{\bibinfo{person}{Wietse Venema}.}
  \bibinfo{year}{2014}\natexlab{}.
\newblock \bibinfo{title}{{Re: Official Postfix source code repository?}}
\newblock \bibinfo{howpublished}{Reply on mailing list mailing.postfix.users}.
\newblock
\urldef\tempurl%
\url{https://groups.google.com/g/mailing.postfix.users/c/6Kkel3J_nv4/m/fFWPVHDM9XUJ}
\showURL{%
\tempurl}
\newblock
\shownote{(accessed: 2024-04-11)}.


\bibitem[Watson(2013)]%
        {historylzma2024}
\bibfield{author}{\bibinfo{person}{Reilly Watson}.}
  \bibinfo{year}{2013}\natexlab{}.
\newblock \bibinfo{title}{{History of LZMA Utils and XZ Utils}}.
\newblock
\newblock
\urldef\tempurl%
\url{https://github.com/kobolabs/liblzma/blob/87b7682ce4b1c849504e2b3641cebaad62aaef87/doc/history.txt}
\showURL{%
\tempurl}
\newblock
\shownote{(accessed: 2024-04-11)}.


\bibitem[Watson et~al\mbox{.}(2015)]%
        {watson2015cheri}
\bibfield{author}{\bibinfo{person}{Robert N.~M. Watson},
  \bibinfo{person}{Jonathan Woodruff}, \bibinfo{person}{Peter~G. Neumann},
  \bibinfo{person}{Simon~W. Moore}, \bibinfo{person}{Jonathan Anderson},
  \bibinfo{person}{David Chisnall}, \bibinfo{person}{Nirav Dave},
  \bibinfo{person}{Brooks Davis}, \bibinfo{person}{Khilan Gudka},
  \bibinfo{person}{Ben Laurie}, {et~al\mbox{.}}}
  \bibinfo{year}{2015}\natexlab{}.
\newblock \showarticletitle{{CHERI: A hybrid capability-system architecture for
  scalable software compartmentalization}}. In \bibinfo{booktitle}{\emph{2015
  IEEE Symposium on Security and Privacy}}. \bibinfo{publisher}{IEEE},
  \bibinfo{pages}{20--37}.
\newblock
\urldef\tempurl%
\url{https://doi.org/10.1109/SP.2015.9}
\showDOI{\tempurl}


\end{thebibliography}

\end{document}